\documentclass[a4paper,10pt]{article}
\usepackage{graphicx,natbib}
\newcommand{\rmn}[1]{{\mathrm{#1}}}
\def\apj{ApJ\,  }
\def\apjl{ApJ\,  }

\def\mnras{MNRAS\,  }
\def\pasj{PASJ\,  }

\def\pla{Phys. Lett. A   }
\def\za{Z. Astrophys.  } 
\title
{
The oscillating behavior of the
pair correlation function  in galaxies
}
\author{L. Zaninetti }

\begin{document}
\maketitle 

Dipartimento di Fisica , Via Pietro Giuria 1, 10125, Turin, Italy

\begin{abstract}
The pair correlation function (PCF) for galaxies presents
typical oscillations in the range 20-200 Mpc/h which are named
baryon acoustic oscillation (BAO).
We first review and test the
oscillations of the PCF when the  2D/3D vertexes
of  the Poissonian Voronoi Tessellation (PVT) are considered.
We then model the behavior  of the PCF at a small scale
in the presence of an auto gravitating  medium
having a line/plane of symmetry in 2D/3D.
The analysis of the PCF in an astrophysical context
was split into two,  adopting  a non-Poissonian
Voronoi Tessellation (NPVT).
We first analyzed the case of a 2D cut which covers
few voids and a 2D cut which covers approximately
50 voids.
The obtained PCF in the  case of many voids
was then discussed
in comparison to the
bootstrap predictions for
a  PVT  process
and the observed PCF for an astronomical catalog.
An approximated formula which connects the averaged radius of the cosmic
voids  to the first minimum of the PCF is given.
\end{abstract}

\section{Introduction}

The computation of pair correlation functions (PCF)
started with the experimental evaluation of the
PCF of liquid argon from neutron scattering,
see \cite{Yarnell1973}.
The PCF was then treated in many  books
as a subject  connected  to mono-atomic fluids,
see \cite{McQuarrie1976,Hansen1986,Allen1987}.
The PCF was later applied to the
Poissonian Voronoi
Tessellation (PVT)
and the PCF for  2D/3D vertexes
was evaluated, see \cite{Stoyan1990,Okabe2000,Heinrich2008}.
The astronomers started to study the  baryon
acoustic oscillations (BAO)
as a tool for probing the cosmological distance
scale and dark energy,
see \cite{Eisenstein2005}.
The astronomical analysis  continued
with the study of:
(i) the  impact of uncertainties in the
photometric redshift error probability
distribution on dark energy constraints
from the detection of BAO  in galaxy power spectra,
see \cite{Zhan2006},
 (ii) the cosmological distance errors achievable
using the BAO as a standard ruler,
see   \cite{Seo2007},
(iii) the detectability  of BAO in the power spectrum
of galaxies using ultra large volume $N$-body
simulations of the hierarchical clustering of dark matter
and semi-analytical modeling of galaxy formation,
see \cite{Angulo2008},
(iv)  a set of ultra-large particle-mesh simulations
of the Lyman-$\alpha$  forest targeted at understanding
the imprint of BAO in the inter-galactic medium,
see \cite{White2010},
(v) the use of BAO  to map the expansion history
of the universe, see \cite{Mehta2011},
(vi) a measure of BAO  from the angular power
spectra of the Sloan Digital Sky Survey
III (SDSS-III) Data Release 8 imaging catalog
that includes 872,921 galaxies over 10000 ${\rmn {deg}}^2$
between 0.45 $< z < 0.65$, see \cite{seo2012}.

\section{The PCF }

This Section outlines the  difference
between PCF in fluids
and PCF in astronomy.
The PCF adopted for Voronoi Diagrams is reviewed.

\subsection{The PCF in fluids}

The PCF  gives the probability of finding
a pair of objects a distance $r$ apart,
relative to the probability
expected for a random distribution of the same density.
This function can be estimated as
\begin{equation}
g(r)=\frac{V}{N^2}
\langle
\Sigma_j \Sigma_{j\neq i} \delta ( { \bf  r} - {\bf r}_{ij} )
\rangle
\quad  ,
\end{equation}
where $V$ is the considered volume,
$N$ is the number of objects, and  ${\bf r}_{ij}$
is the  distance between centers,
see formula (2.94) in  \cite{Allen1987}.

\subsection{The PCF in astronomy}

Astronomers, beginning with \cite{Totsuji1969} ,
have used  the following convention
for the PCF  for galaxies
\begin{equation}
\xi_{\rmn{GG}}(r) = 1 +g(r)
\quad .
\end{equation}
The  most used conventions for low values of
$r$
assumes that
\begin{equation}
 \xi _{\rmn{GG}} = ({r \over r_{\rmn {G}}})^{-\gamma_{\rmn{GG} }}
\quad ,
\label{parameterscorre}
\end{equation}
where  $\gamma_{\rmn{GG} }$=1.8  and
$r_{\rmn {G}} = 5.77h^{-1} {\rmn {Mpc}} $
(the correlation length) when the range
$0.1 h^{-1} {\rmn {Mpc}}  < r < 16 h^{-1} {\rmn {Mpc}} $ is considered,
see  \cite{Zehavi_2004}
where 118,149 galaxies were
analyzed.
Another   estimator  of the PCF
useful both for simulations of vertexes of PVT
and catalogs of galaxies
is
\begin{equation}
 \xi_{\rmn {OO}} = 1
+ \frac { n_{\rmn {OO}}(r) } {n_{\rmn {RR}}(r)}
- 2  \frac { n_{\rmn {OR}}(r) } {n_{\rmn {RR}}(r)}
 \quad .
 \label {csimio}
\end {equation}
where
$n_{\rmn {OO}}(r)$,
$n_{\rmn {RR}}(r)$,  and
$n_{\rmn {OR}}(r)$
are the number of
object--object,
random--random,
and
object--random
pairs having distance $r$,
see \cite{Szalay1993,Martinez2009};
in this paper  the word object can be substituted
by a vertex of a PVT or a galaxy.
The random configuration can be realized
every time with a different
random number generator.
As an example, we generate $N_{\rmn {T}} $ different random configurations
and we average the different values of
$\xi_{\rmn {OO}}$ which  now
will be characterized by
an error bar given by the standard deviation.

\subsection{The PCF in Voronoi Diagrams}

\begin{figure*}
\begin{center}
\includegraphics[width=10cm]{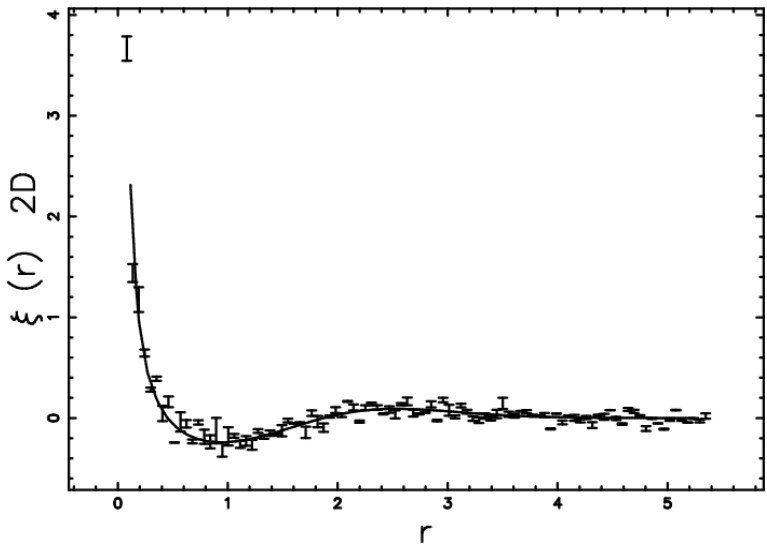}
\end {center}
\caption
{
The values   of the PCF  for the point process of
vertexes of 2D PVT, in ${\bf R}^2$ (full line)
as extracted from Table 5.6.1  in Okabe et al. (2000).
The simulated PCF  as given by Eq. (\ref{csimio}) is
reported as the dashed line with error bar.
The distances are expressed in normalized units,
the number $N_{\rmn{ s}}$ of seeds is 1000,
and $N_{\rmn {T}} $=10.
}
\label{f01}
    \end{figure*}

\begin{figure*}
\begin{center}
\includegraphics[width=10cm]{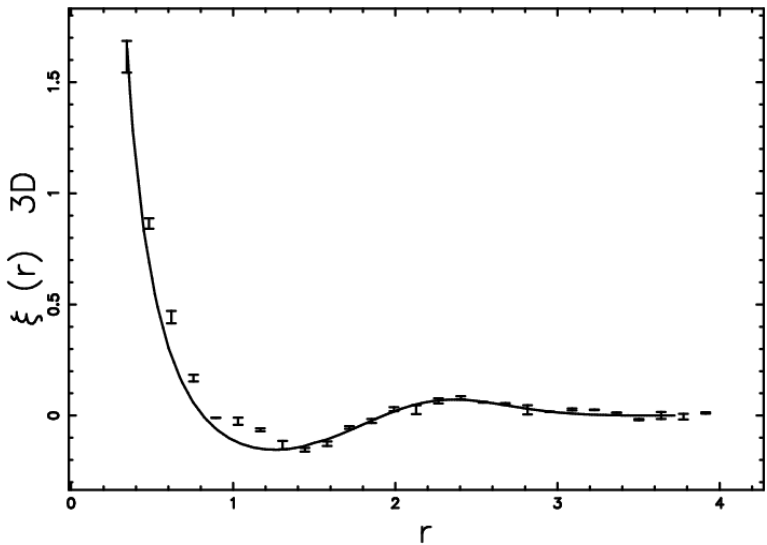}
\end {center}
\caption
{
Values of the PCF  for the point process of
vertexes of 3D PVT, in ${\bf R}^3$ (full line)
as extracted from Table 5.6.2  in Okabe et al. (2000).
The simulated PCF  as given by Eq. (\ref{csimio}) is
reported as a dashed line with error bar.
The distances are expressed in normalized units,
the number, $N_{\rmn{ s}}$, of seeds is 800 and  $N_{\rmn {T}} $= 5.
}
\label{f02}
    \end{figure*}

The  PCF  for the point
process of vertexes in
${\bf  R}^2$ and ${\bf R}^3$
has  been analyzed  by \cite{Stoyan1990,Okabe2000,Heinrich2008}.
A review formula for the PCF is  reported in \cite{Okabe2000},
see Eqn. (5.6.2)
with m=2 (2D) and m=3 (3D) and Tables 5.6.1 and 5.6.2;
as a first analysis we tested
the  2D and 3D results with our simulation.
In this case  the PCF is computed  via the numerical
formula (\ref{csimio}).
The adopted scale is important  and in view of
future applications we have chosen
the generalized  averaged radius, $\bar{R}$,
of the 2D cells  as
\begin{equation}
\pi {\bar{R}}^2 = \frac{L^2}{N}
\quad ,
\end{equation}
where    $L$ is the side of the square
in which $N_{\rmn{ s}}$ seeds are inserted.
The generalized averaged radius in 3D
is
\begin{equation}
\frac{4}{3} \pi {\bar{R}}^3 = \frac{L^3}{N}
\quad ,
\end{equation}
where    $L$ is the side of the cube in which
         $N_{\rmn{ s}}$ seeds are inserted.
The distances will therefore be expressed in normalized
units $r/\bar{R}$
and as an example a normalized distance of 2
corresponds to an averaged diameter.
Figures (\ref{f01}) and (\ref{f02})
report the 2D/3D mathematical
PCF as well our simulated one in normalized units.

We are in  the presence of
a damped  oscillatory
behavior of  the PCF   for PVT vertexes
and
the first minimum is reached  when
$r/\bar{R} \approx 1 $ in 2D
and  $r/\bar{R} \approx 1.4 $ in 3D.

\section{Galaxies on the faces}

This section first reviews the statistical distributions
as function of the height  from a plane  of reference
for a system of auto gravitating  galaxies
and then provides a first astrophysical
environment when we  have a symmetrical configuration
of galaxies in respect to a line (2D) or
in respect  to a plane (3D).

\subsection{Self-gravitating  galaxies}

\begin{figure*}
\begin{center}
\includegraphics[width=10cm]{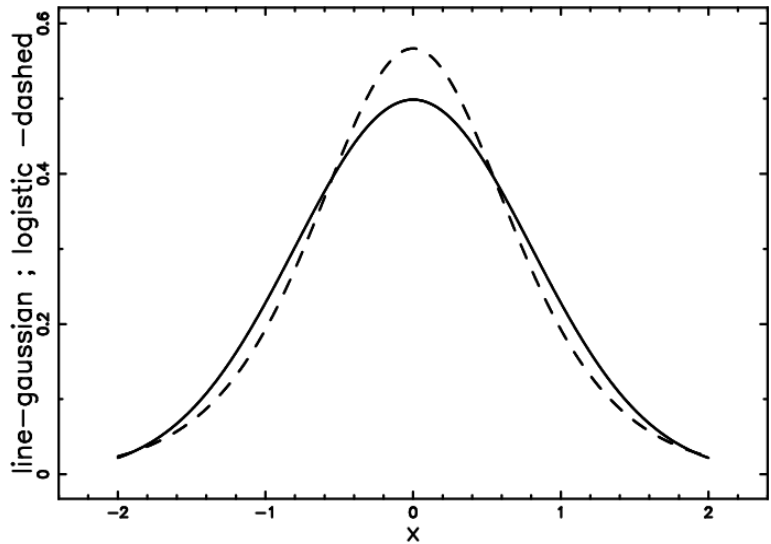}
\end {center}
\caption
{
Normal PDF (full line) and
logistic PDF  as
represented by
Eq. (\ref {sech2variance})
(dashed line)
when $\sigma=0.8 {\rmn {Mpc}}/h $.
}
\label{f03}
    \end{figure*}

The density profile of a thin
self-gravitating disk of gas
which is characterized by a
Maxwellian distribution in velocity
and  a distribution which varies
only in the $z$-direction is here applied to galaxies,
\begin{equation}
n(z;z_0) = n_0 sech^2 (\frac{z}{2\,z_0})
\quad ,
\label{sech2}
\end{equation}
where
$n_0$ is the galaxy density at $z=0$,
$z_0$ is a scaling parameter in     ${\rmn {Mpc}}/h $,
and  $sech$ is the hyperbolic secant
(\cite{Spitzer1942,Rohlfs1977,Bertin2000,Padmanabhan_III_2002}).
This physical law can be converted  to a
probability density function (PDF), the probability of
having a galaxy at a distance
between $z$ and $z+dz$ from a plane
of symmetry
\begin{equation}
p(z;z_0)=
\frac{1}{4}\, \left( {\it sech} \left(\frac{ 1}{2}\,{\frac { \left| z \right| }{z_{{0}
}}} \right)  \right) ^{2}\frac{1}{z_0}
\quad .
\label{sech2prob}
\end{equation}
The range of existence of
this PDF, which is the logistic distribution,
is in the interval $[-\infty, \infty]$,
see \cite{Balakrishnan1991handbook,univariate2,evans}.
The average value is
$E(z;z_0) =0$ and the variance is
\begin{equation}
\sigma^2(z;z_0)=
\frac{1}{3}\,{z_{{0}}}^{2}{\pi }^{2}
\quad .
\end{equation}
This PDF can be converted in such a way that
it can be compared with the
normal (Gaussian)
distribution
\begin{equation}
p_{\rmn {N}} (z;\sigma) =
\frac {1} {\sigma (2 \pi)^{1/2}}  \exp {- {\frac {z^2}{2\sigma^2}}}
\quad  ,
\label{gaussian}
\end{equation}
where $z$ is the distance in ${\rmn {Mpc}}/h $ from the plane
of symmetry and $\sigma$ the standard deviation in ${\rmn {Mpc}}/h $.
The substitution $z_0={\frac {\sqrt {3}\sigma}{\pi }}$
transforms the PDF
(\ref{sech2prob})  into
\begin{equation}
p(z;\sigma)=
\frac{1}{12}\, \left( {\it sech} \left( \frac{1}{6}\,{\frac { \left| z \right| \pi \,
\sqrt {3}}{\sigma}} \right)  \right) ^{2}\pi \,\sqrt {3} \frac {1}{\sigma}
\quad,
\label{sech2variance}
\end{equation}
which has variance $\sigma^2$.
The similarity with the normal distribution is
straightforward and Figure \ref{f03} reports
the two PDFs when  the value of
$\sigma$ is equal in both cases.

\subsection{2D and 3D symmetry }

\begin{figure*}
\begin{center}
\includegraphics[width=10cm]{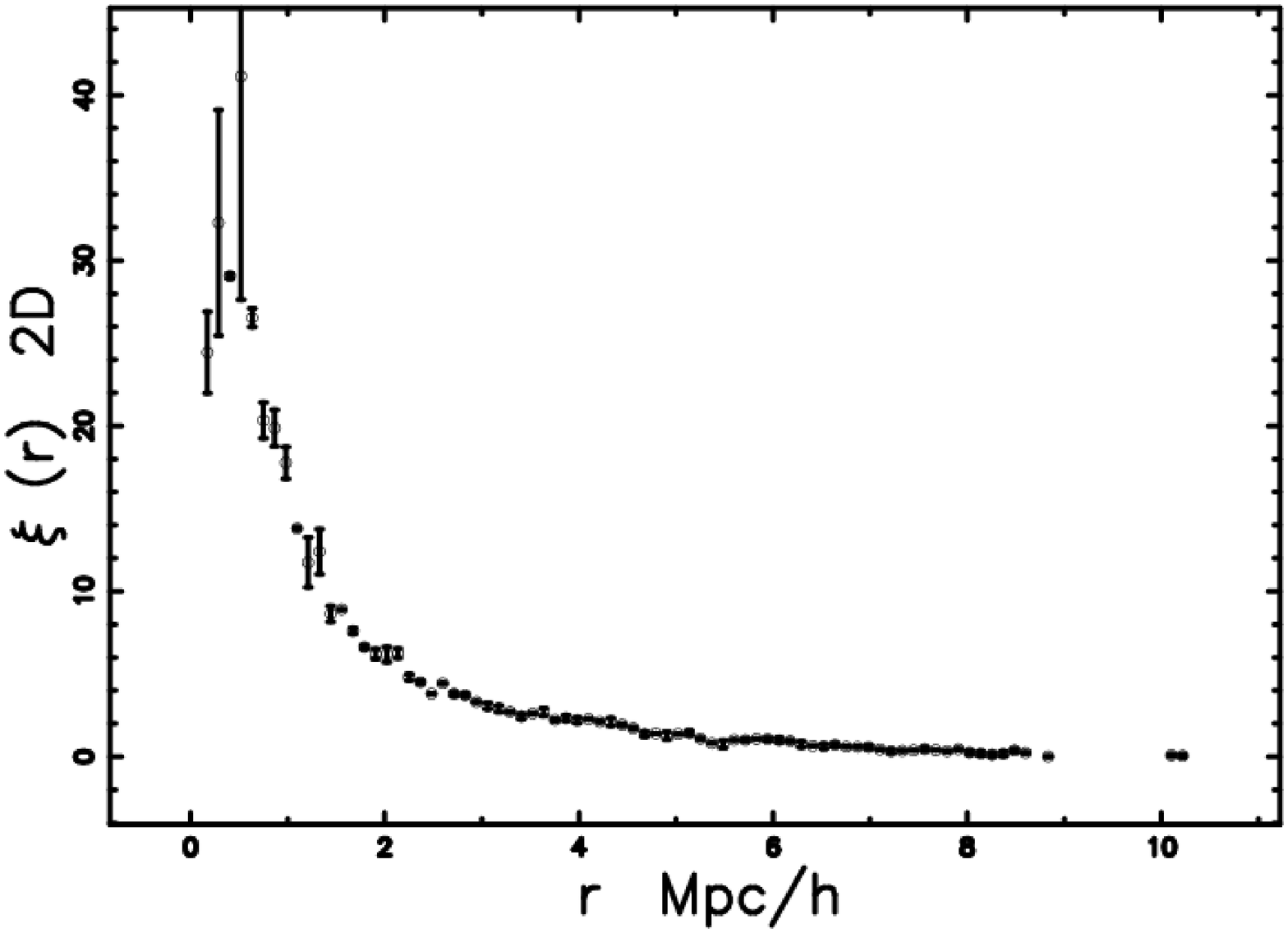}
\end {center}
\caption
{
Values of the PCF
for    200 galaxies
generated according
to  the
logistic PDF as represented
by Eq. (\ref {sech2variance})
in the $z$-direction;
$\sigma=0.4 {\rmn {Mpc}} /h$,
the length of the line of symmetry
is 40 ${\rmn {Mpc}}/h $ and  $N_{\rmn {T}} $= 20.
The PCF as approximated by Eq. (\ref{parameterscorre})
has  parameters
$\gamma_{\rmn{GG} }$=1.52 and  $r_{\rmn {G}} = 5.27 h ^{-1} {\rmn {Mpc}} $.
}
\label{f04}
    \end{figure*}

\begin{figure*}
\begin{center}
\includegraphics[width=10cm]{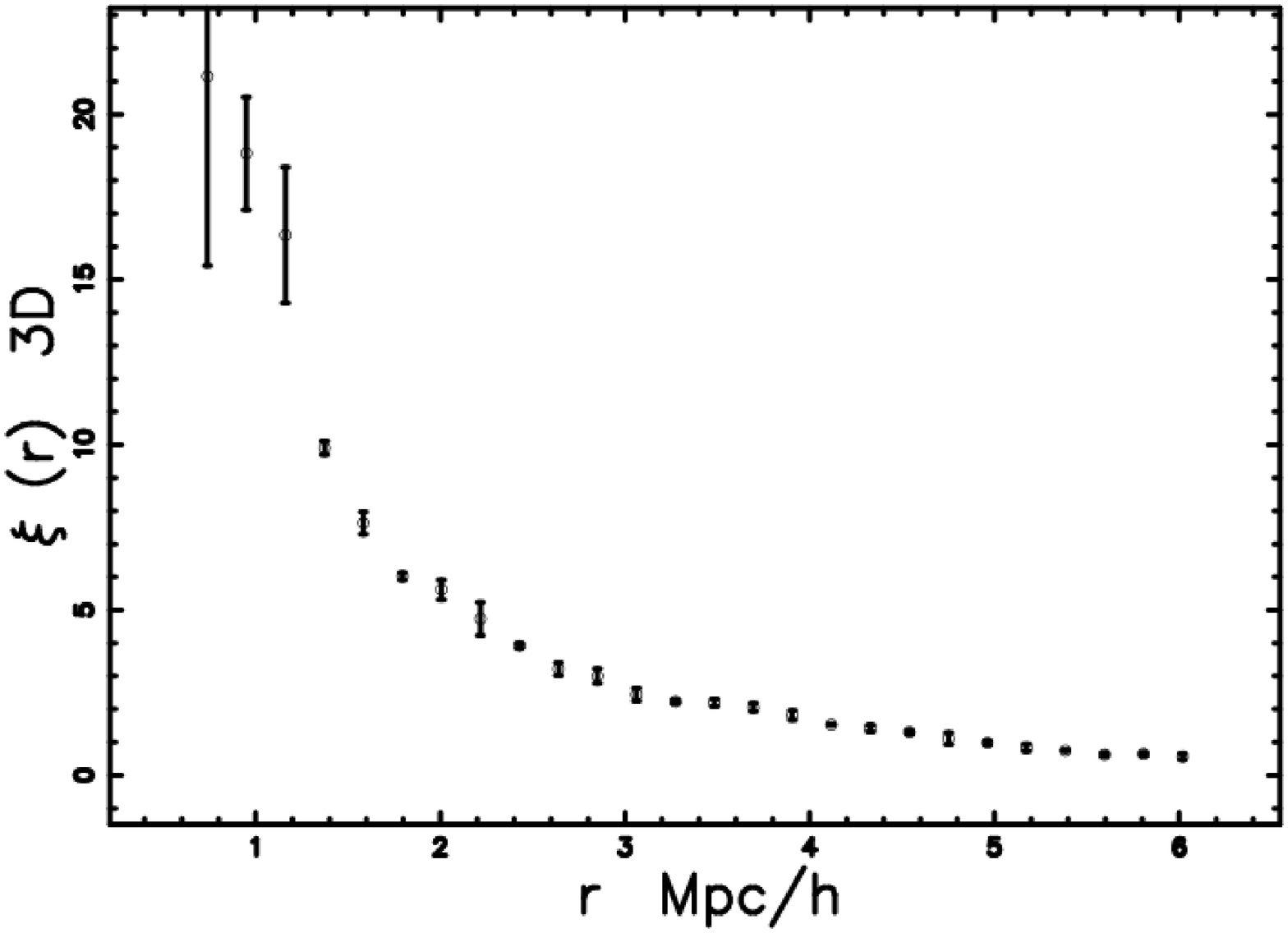}
\end {center}
\caption
{
Values of the PCF
for  200 galaxies  generated according to
to  the
logistic PDF as represented by
Eq. (\ref {sech2variance})
in the $z$-direction;
$\sigma=0.1 {\rmn {Mpc}}/h  $,
the side of the plane of symmetry
is 20 ${\rmn {Mpc}}/h $
and $N_{\rmn {T}} $= 30.
The PCF as approximated by Eq. (\ref{parameterscorre})
has  parameters
$\gamma_{\rmn{GG} }$=1.8  and  $r_{\rmn {G}} = 4.97 h ^{-1} {\rmn {Mpc}} $.
}
\label{f05}
    \end{figure*}

We now analyze the astrophysical  case in which
the galaxies follow the logistic law along a
perpendicular direction, $z$-axis,
to a line along the      $y$-axis (2D)
and are randomly disposed in the $y$ direction.
By analogy with the 3D  case  the galaxies
will follow the logistic law in
the    $z$-direction and will be randomly generated
in the $x$ and $y$ directions.
In other words we realize a statistical
equilibrium for galaxies
with respect to a plane or to a line.
Figures \ref{f04} and \ref{f05} report the
computation of the 2D/3D PCF with a choice
of parameters which
give values of $\gamma_{\rmn{GG} }$   and  $r_{\rmn {G}} $
similar to those  observed, see Eq. (\ref{parameterscorre}).

\section{Astrophysical Applications}

Up to now we have expressed all the distances
in normalized units in which the unit is the averaged
radius of the approximated circles/spheres which
approximate the 2D/3D  area/volume  of a Voronoi cell.
A second unit system is obtained multiplying the normalized
distance $r_u$ by   the  averaged radius of  the voids
,$\overline{R}$ , or  $r=r_u\overline{R}$ ; this can be considered the physical   space.
A third unit system is connected with the redshift, $z$,
 which after
\citet{Hubble1929}
\begin{equation}
z = \frac{H_0 D} {c}
\quad ,
\end{equation}
where
 $H_0 = 100 h \mathrm{\ km\ s}^{-1} \mathrm{\ Mpc}^{-1}$ , with $h=1$
when  $h$ is not specified,
$D$ is the distance in $Mpc$  and
$c$ is  the  light velocity.
In our framework
\begin{equation}
z = \frac{H_0 \, r_u\overline{R}} {c}
\quad
\end{equation}
and this conversion defines the redshift space.
Voronoi diagrams can model the local universe
once  two different calibrations are done.
The first calibration is connected with
the averaged effective radius  of the voids.
A first analysis of the Sloan Digital Sky Survey (SDSS)
R7 catalog, see \cite{Vogeley2011},
suggests
$\overline{R} = 18.23 \,{\rmn {Mpc}}/h$
as  the averaged effective radius  of the voids.
 A second  analysis of  the same catalog finds
 that the effective radii of the voids  range from 5 to 135  Mpc/h
 and a first approximated  evaluation gives
 $\overline{R} = 70 \,{\rmn {Mpc}}/h$ , see \citet{Sutter2012}.
 We therefore define an acceptable range of variability for the
 averaged effective radius  of the voids
 \begin{equation}
 16.23 \,{\rmn {Mpc}}/h \leq \overline{R} \leq 70 \,{\rmn {Mpc}}/h
 \quad .
 \end{equation}
The  theoretical averaged  radius  of the Voronoi
volumes approximated
by  spheres   should  be comprised in this current range of variability;
this is  the first  calibration.
On adopting a practical    point of view
$\overline{R}$  will be fixed in such a way
that the first minimum  of the PCF is at  65 Mpc/h.
A second  calibration  of the Voronoi diagrams
originates from the PDF which models
the statistics of the voids.
A possible  model  for the  observed statistics
of the voids' volumes
as given by  SDSS R7
is the Kiang function,
see \cite{kiang},
\begin{equation}
 H (x ;c ) = \frac {c} {\Gamma (c)} (cx )^{c-1} \exp(-cx),
\label{kiang}
\end{equation}
with  $c=2$, see \cite{Zaninetti2012e}.
Due  to the fact  that  the PVT  seeds are characterized
by
$c=5$ in the Kiang function we generate $N_{\rmn {s}}$ seeds,
in order
to have $c=2$.
see  \cite{Zaninetti2013b}.
The points of a tessellation in 3D are of four types,
depending
on how many nearest
neighbors in $ES$, the ensemble of seeds, they have.
The name seeds  derive  from their  role in generating cells.
Basically  we have two kinds  of seeds , Poissonian and non
Poissonian  which generate the Poissonian Voronoi tessellation
(PVT) and the non Poissonian Voronoi tessellation (NPVT). The
Poissonian   seeds  are generated independently on the $X$, $Y$
and $Z$ axis in 3D through a subroutine  which returns a
pseudo-random real number taken from a uniform distribution
between 0 and 1. 
This  is the case most studied and for 
practical
purposes,
the subroutine
RAN2  was used, see \cite{press}.
The non Poissonian   seeds  can  be  generated
in an infinite  number  of different ways:
some examples of  NPVT  are reported  in  \cite{Zaninetti2009c},
here the seeds  are  generated in order that 
the PDF in volumes follows
a Kiang function with $c=2$.
The  algorithm that  generates such seeds is reported 
in  Section 3.3 of  \cite{Zaninetti2013b}. 
A point with exactly one nearest neighbor's is in
the interior of a cell,
a point with two nearest neighbors is on the face between
two cells,
a point with
three nearest neighbors is
on an edge shared by three cells,
and a point with four neighbors is a
vertex where three cells meet.
Following the nomenclature introduced
by \cite{okabe}, 
we call the intersection between a plane and
the PVT 
$V_p(2,3)$ and  $V_{np} (2,3)$ 
the intersection between a plane and
the NPVT.
A discussion  on how to extract  the edges in 3D or the 
faces that in  $V_p(2,3)$ and
$V_{np} (2,3)$ become lines,  can be found in 
Section 4.1 of \cite{Zaninetti2010a}. 
We  now explore two cases: the first case is a  2D cut
 or $V_{np} (2,3)$  
on
a 3D network  which  covers few voids
and the  second one is a cut  which covers many voids.

\subsection{Few  voids}
\begin{figure*}
\begin{center}
\includegraphics[width=10cm]{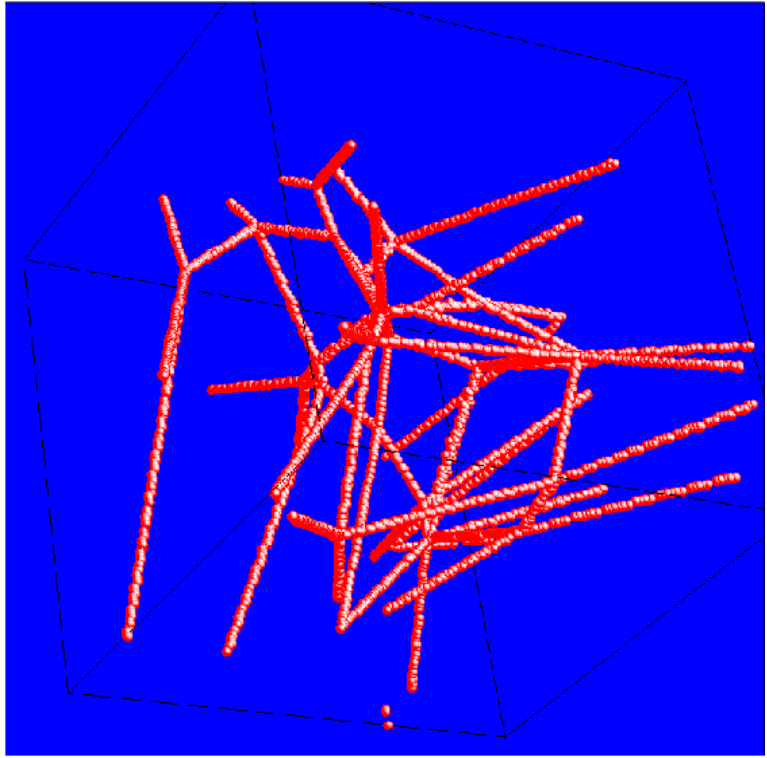}
\end {center}
\caption{
3D visualization of the
edges of few voids in NPVT.
The  parameters
are      $ N_{\rmn {s}}   $   = 20,
         $ side  $   = 213 ${\rmn {Mpc}} /h$ and
         $ h=1$.}
          \label{f06}%
    \end{figure*}

The 3D  network  of few voids 
can be visualized
through  the display  of the edges,
see Figure \ref{f06}.

\begin{figure*}
\begin{center}
\includegraphics[width=10cm]{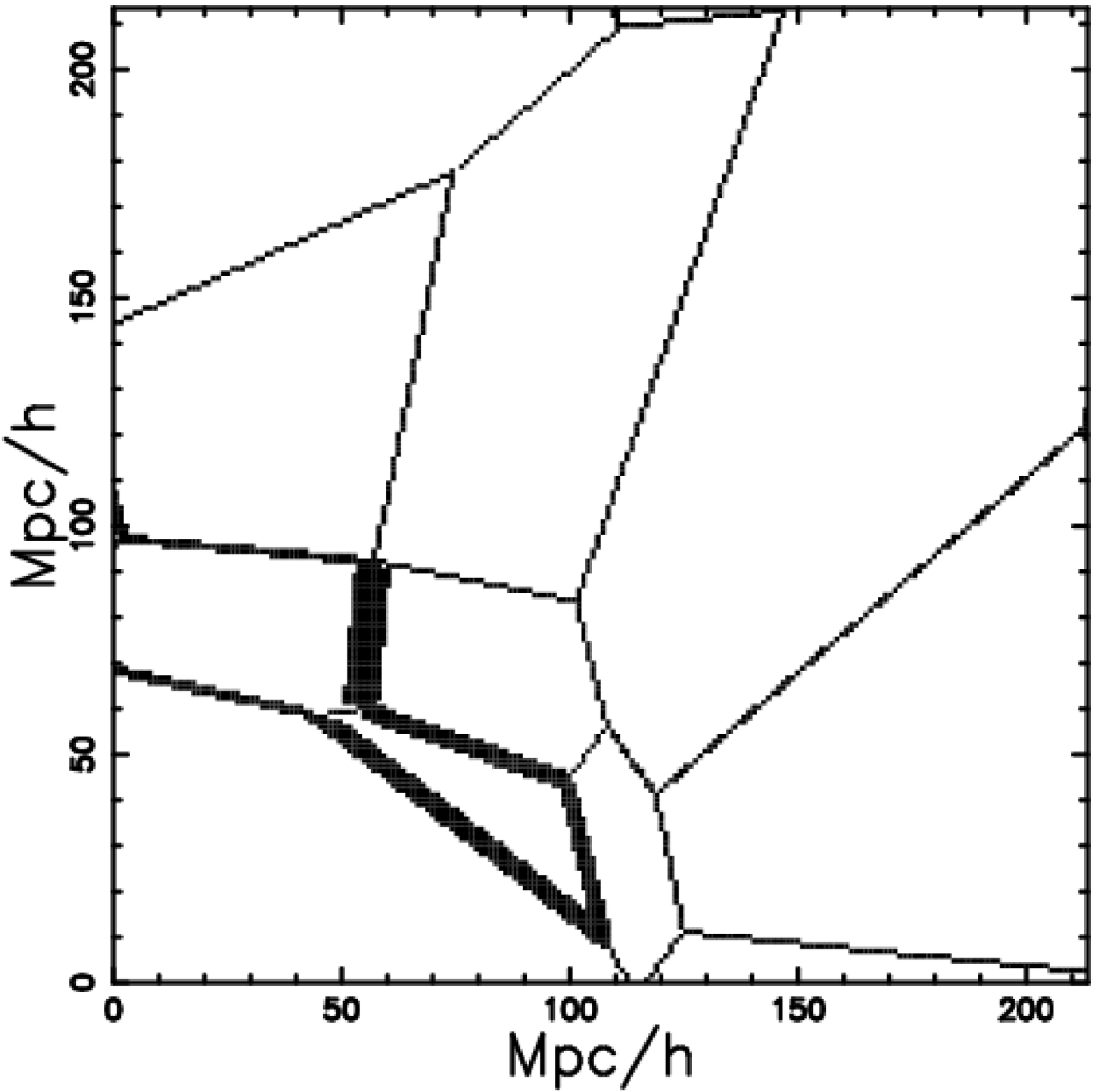}
\end {center}
\caption
{
Cut or $V_{np} (2,3)$ on the  $X-Y$ plane
in the middle of the box
of the NPVT faces.
}
\label{f07}%
\end{figure*}

We now consider  a 2D cut
or $V_{np} (2,3)$
  on the  3D NPVT network
of faces, see  Figure \ref{f07}.

\begin{figure*}
\begin{center}
\includegraphics[width=10cm]{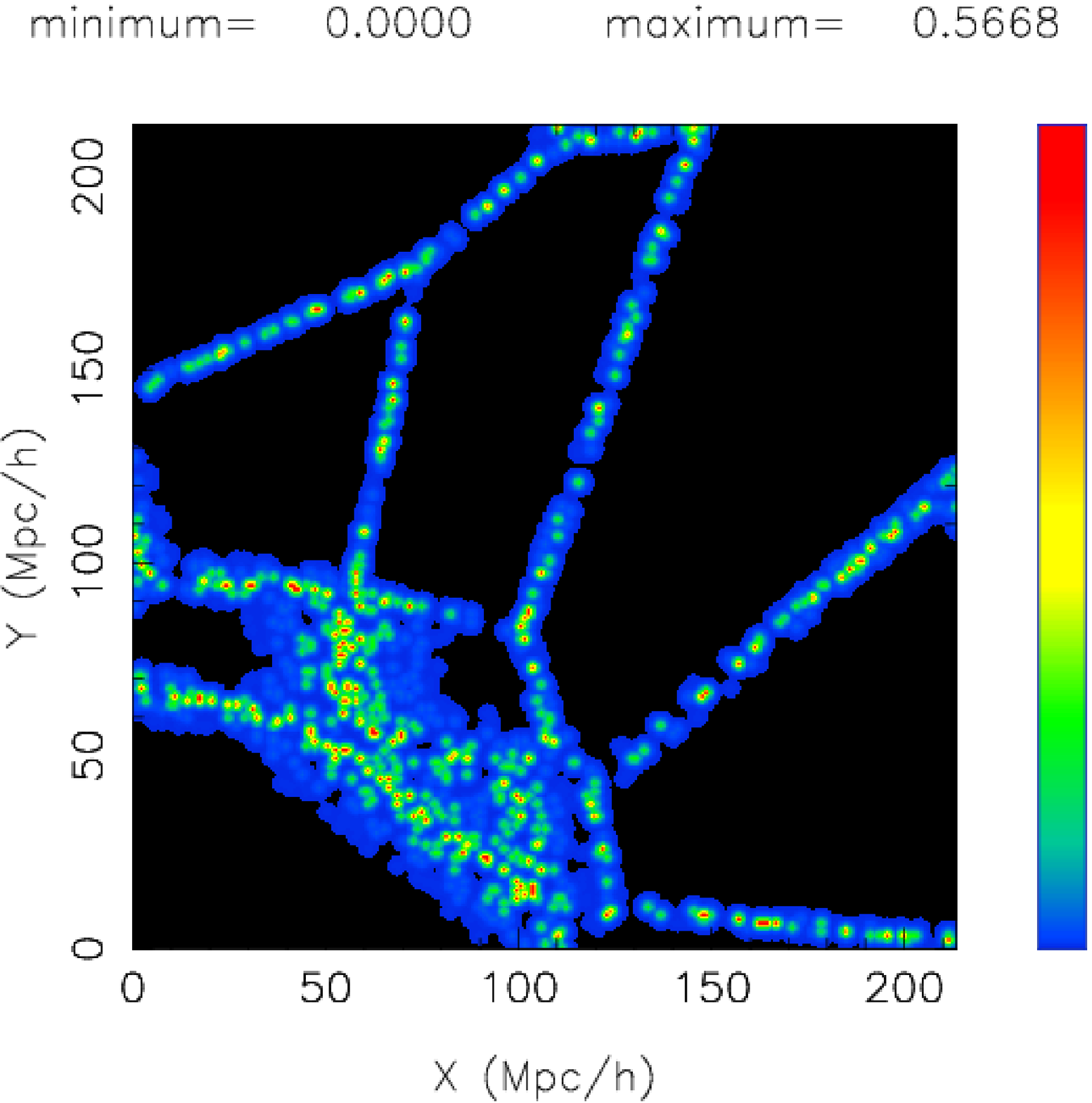}
\end {center}
\caption{
2D map
of the probability of having
a galaxy.
The  Voronoi parameters are the same as in
Figure~\ref{f06} and $\sigma=0.8 {\rmn {Mpc}}/h $.
The $X$ and $Y$ units are in ${\rmn {Mpc}} /h $.
        }
    \label{f08}
    \end{figure*}

\begin{figure*}
\begin{center}
\includegraphics[width=10cm]{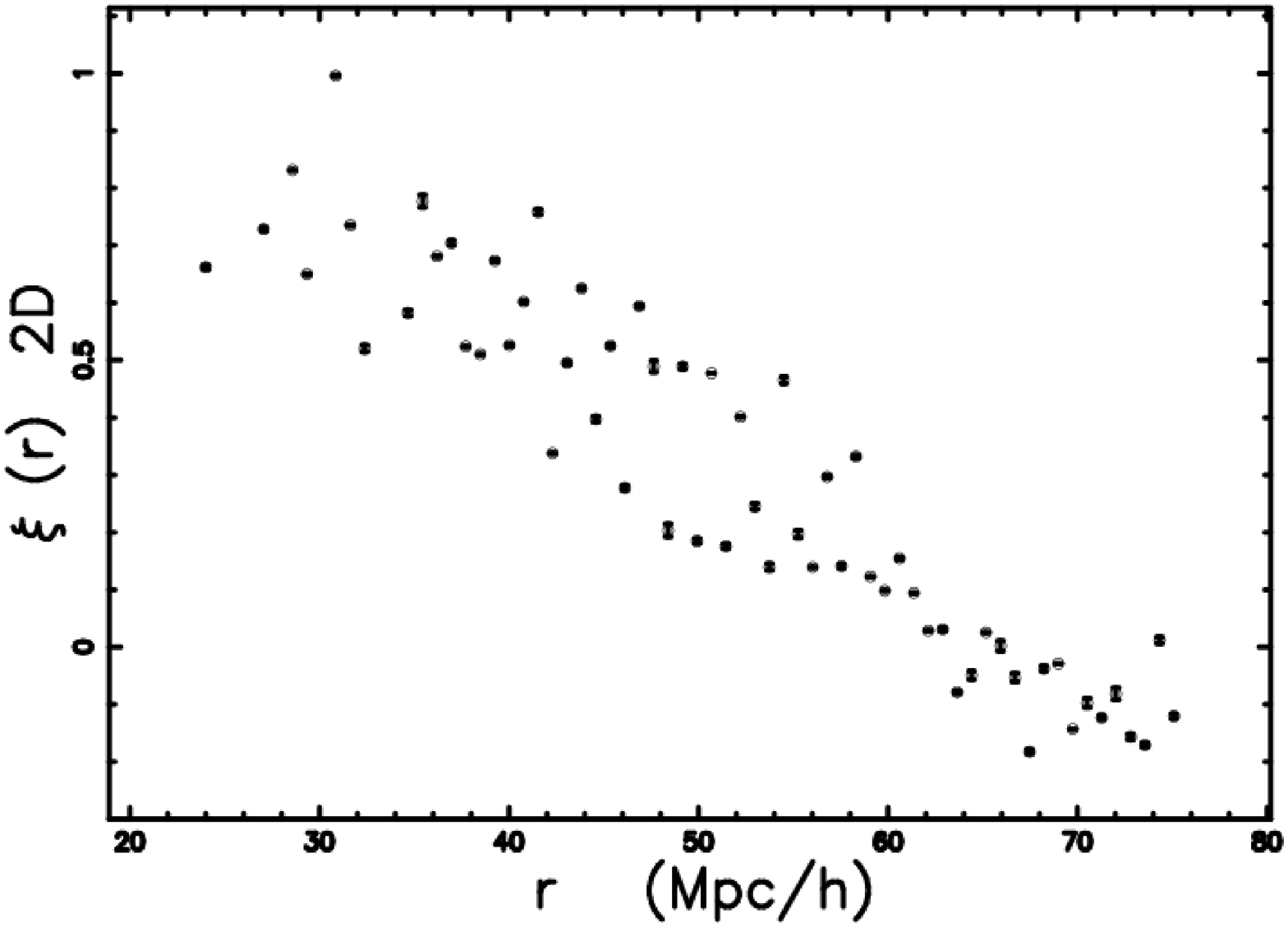}
\end {center}
\caption
{
Values of the PCF
for    10,000 galaxies
generated according to
the
logistic PDF,
Eq. (\ref {sech2variance}),
when $\sigma=0.8 {\rmn {Mpc}}/h $.
Here the  $z$ direction is  perpendicular to
the line  which represents the face in a 2D cut
 or $V_{np} (2,3)$  
.
The distances are expressed in Mpc ,
the number of trials is  10  and
the PCF is evaluated using
Eq. (\ref{csimio}).
The position of the first minimum is at 67 Mpc/h.
}
\label{f09}
    \end{figure*}

\begin{figure*}
\begin{center}
\includegraphics[width=10cm]{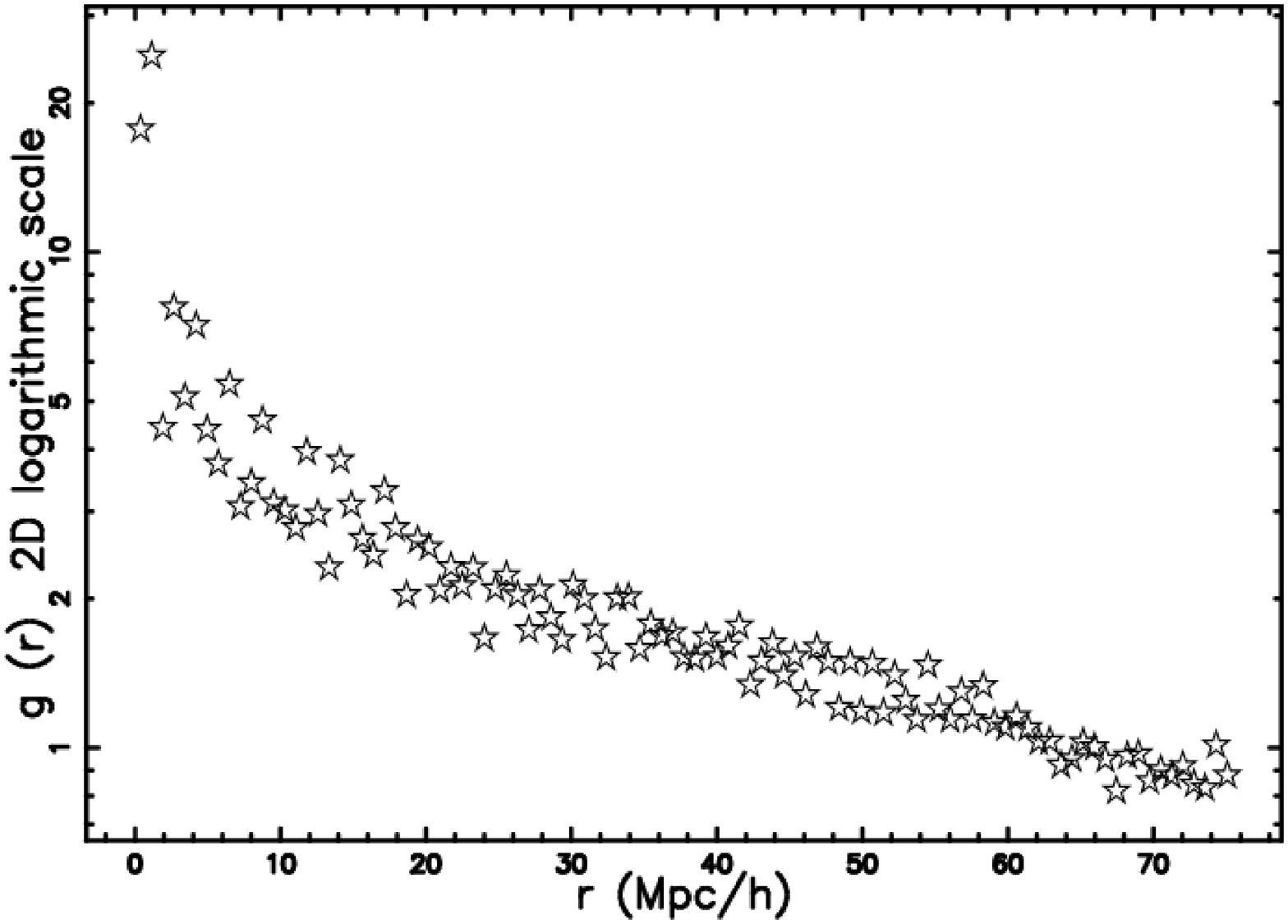}
\end {center}
\caption
{
The same as Figure \ref{f09}
but the vertical  axis has a  logarithmic scale.
}
\label{f10}
    \end{figure*}

We now compute  the probability   of having a galaxy
according  to  the logistic
law,
see Eq. (\ref{sech2variance}) , in
the previous plane where $z$ is the
distance from a face which
in the 2D cut  or $V_{np} (2,3)$   is a line .
Figure \ref{f08}
reports  the values of the probability of having a galaxy
in  a 2D cut  or $V_{np} (2,3)$   organized as  a contour plot.
Figure \ref{f09}
displays the PCF for large values of $r$
and Figure \ref{f10}
the PCF for the full  range  of $r$; the first minimum exists but is not 
well defined.

\subsection{Large Scale Structures}

\label{seclarge}
\begin{figure*}
\begin{center}
\includegraphics[width=10cm]{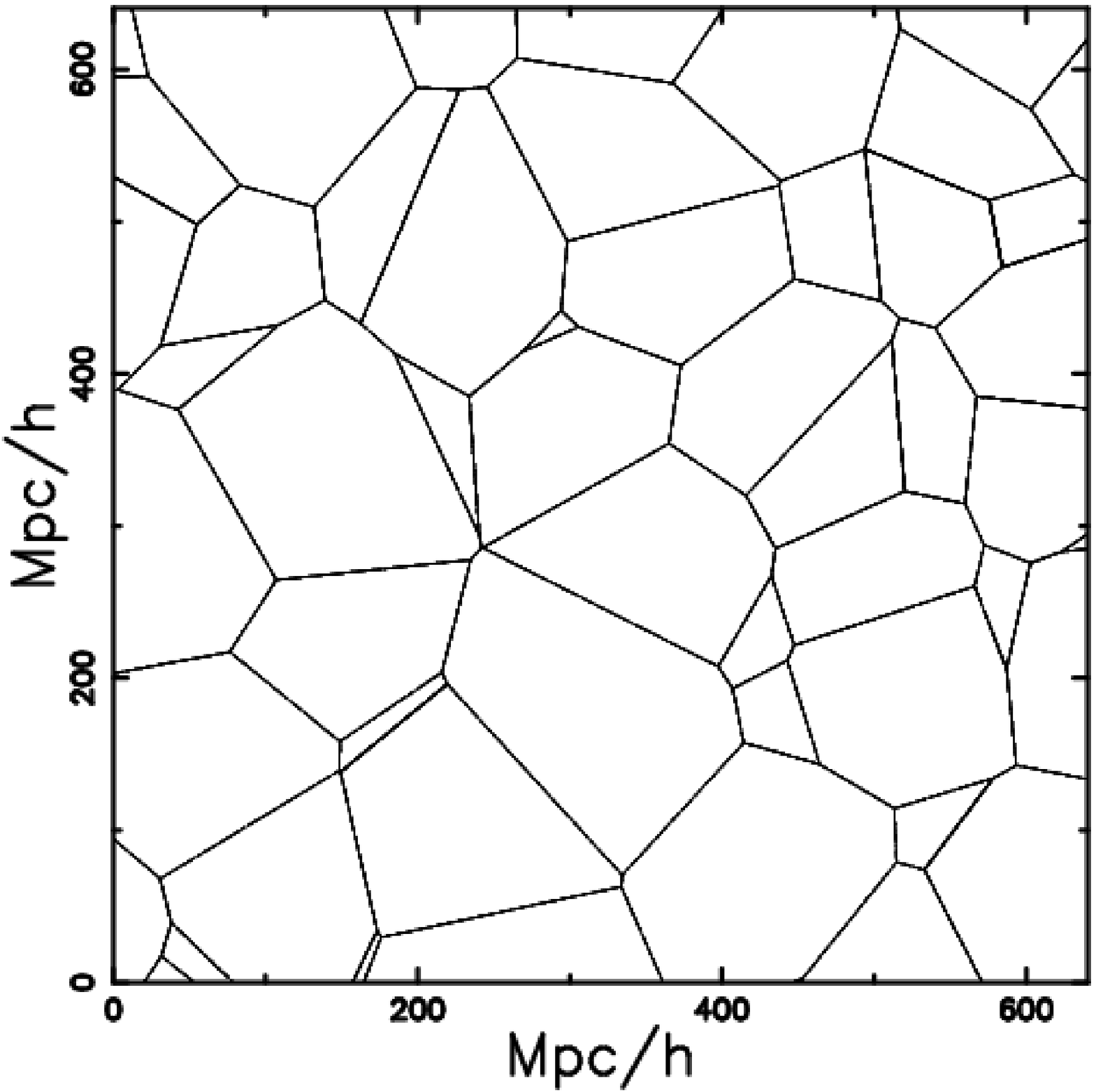}
\end {center}
\caption
{
Network of 30,000 galaxies
selected  on the  $X-Y$ plane
in the middle of the box
which crosses the NPVT faces.
}
\label{f11}%
\end{figure*}

\begin{figure*}
\begin{center}
\includegraphics[width=10cm]{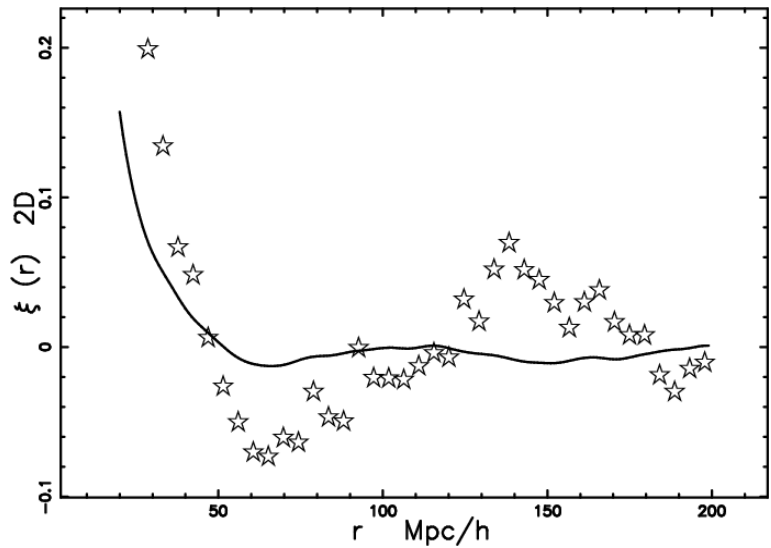}
\end {center}
\caption{
PCF  for the 30,000 selected galaxies (empty stars)
 and the astronomical results  of Martinez et al. (2009)
(full line)
for 2dFVL.
The covered range is
[0-200] Mpc/h
and $N_{\rmn {T}} $ =10.       }
    \label{f12}
    \end{figure*}

We start with a 3D box of $\approx$
600 ${\rmn {Mpc}}/h$ a side
and we produce a 2D
cut which is made by irregular polygons.
On this 2D backbone  we select
30,000 galaxies, see Figure~\ref{f11}.
We then compute the 2D PCF and the results is reported
in Figure \ref{f12} together with the results
of the PCF for 2dFVL,
which is  a nearly VL sample from
2dF Galaxy Redshift Survey (2dFGRS),
see  \cite{Croton2004}.
Our  local minimum is at  65.17  Mpc/h  and
that of Figure 3 (red line) in   \cite{Martinez2009}
at  65 Mpc/h.
\section{Discussion}

\begin{figure}
\begin{center}
\includegraphics[width=10cm]{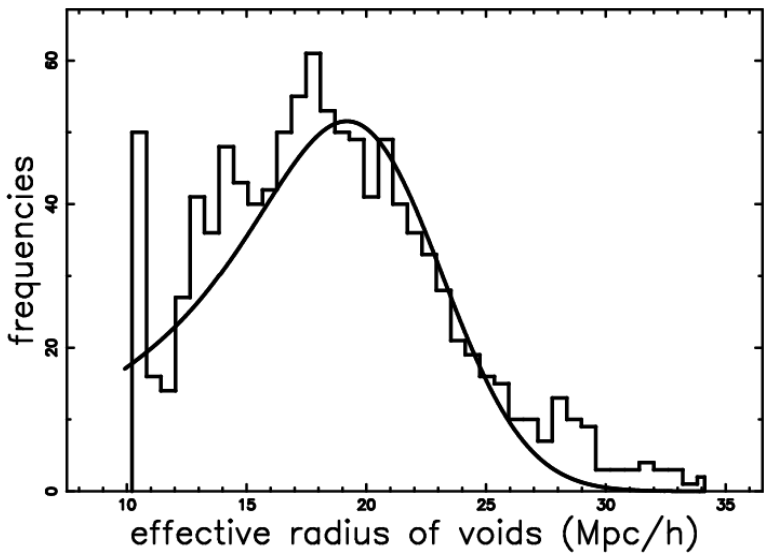}
\end {center}
\caption
{
Histogram (step-diagram)  of
the effective radius of  SDSS DR7
with a superposition of the
PDF  in radius of the PVT 2D cells
as represented by Eq.~(\ref{frmeijer}).
The parameters are
$n$=40, $k$=2,$b$= 34.7/h Mpc ,
$\chi^2$=98 and  $\chi_{red}^2$=2.6.
}
\label{f13}
    \end{figure}

We now outline the differences between  our approach
and the usual model adopted by the astronomers.
The model here adopted is based on the PCF
of the vertexes of PVT
which is developed in a 2D/3D  Euclidean space.
The standard approach  to BAO conversely uses one of the
14 cosmological models available  at the moment , see
\cite{Sanchez2013}.
In the standard approach the first minimum going from the
left to the right  is reached at  $z \approx 0.02 \approx 65 Mpc/h $.
This observational fact allows us to derive
$\overline{R}$ in our model.
The same approximations  are made in the
void finding algorithms applied to
galaxy survey data, which often use  the spheres approximation,
see  \cite{Sutter2012,Nadathur2013,Sutter2013}.
This  approximation is valid when
the thickness  of the surface
which contains the galaxies , $\sigma$ according to eqn.
(\ref{sech2variance}), is  smaller than $\overline{R}$.
A theoretical explanation for  the   variability
of the averaged radius  of the cosmic voids is
perhaps due to the absence of analytical results
for the $V_p(2,3)$ problem.
A new analytical   result for the radius  of the
cell in $V_p(2,3)$ in the PVT case
is
\begin{eqnarray}
f(r,b)=
A
G^{4, 1}_{3, 5}\left({\frac {100}{9}} {\frac {{\pi
}^{2}{r}^{6}}{{b}^{6}}}  \Big\vert ^{5/6, 1/6, 1/2}_{7/3, 2/3, 1/3, 0,
{\frac {17}{6}}}\right) (\frac{1}{b})^{2} \\
 \quad 0 \leq r  \leq b, & ~
\nonumber
\label{frmeijer}
\end{eqnarray}
where
\begin{equation}
A=\frac{2}{3}
{\it K} \sqrt [6]{3}\sqrt [3]{10}\sqrt [3]{\pi }r
\quad ,
\end{equation}
${K}       $ is a constant,
      $K = 1.6485$,
      $r$ is the radius  of the cell,
      $b$ the scale
and  the Meijer $G$-function  is defined
as  in  \citet{Meijer1936,Meijer1941,NIST2010};
see  \cite{Zaninetti2012e} for more details.
Figure \ref{f13}
reports the statistics
of the radius of the cells
$V_p(2,3)$ for the PVT case
as well the frequencies of the
effective radius of  SDSS DR7.
In the previous figure the $\chi^2$
is computed
according to the formula
\begin{equation}
\chi^2 = \sum_{i=1}^n \frac { (T_i - O_i)^2} {T_i},
\label{chisquare}
\end {equation}
where $n  $   is the number of bins,
      $T_i$   is the theoretical  value,
and   $O_i$   is the experimental value
represented
by the frequencies.
The  merit function $\chi_{red}^2$
is  evaluated  by
\begin{equation}
\chi_{red}^2 = \chi^2/NF
\quad,
\label{chisquarereduced}
\end{equation}
where $NF=n-k$ is the number of degrees  of freedom,
      $n$ is the number of bins,
and   $k$ is the number of parameters, in our case 2.

\begin{figure*}
\begin{center}
\includegraphics[width=10cm]{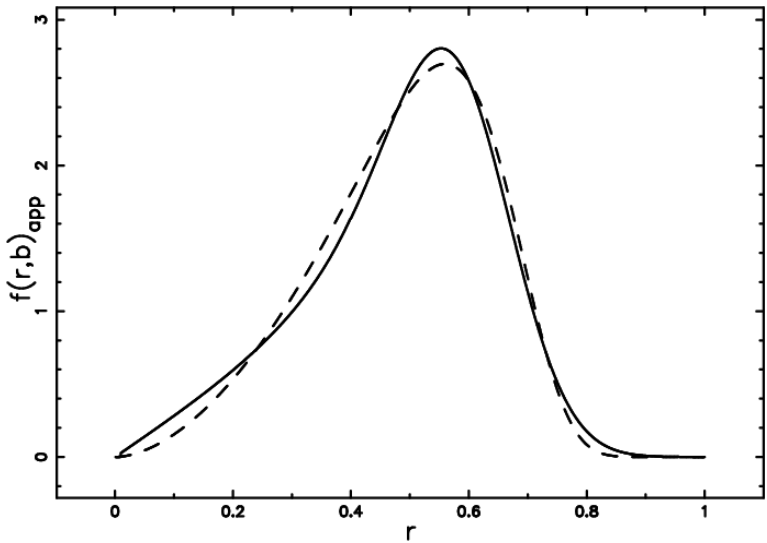}
\end {center}
\caption
{
The analytical  PDF, $f(r,b)$, (full line) and
and  the approximate $f(r,b)_{app}$  
(dashed line).
}
\label{f14}
    \end{figure*}

The averaged effective  radius is  $\overline {r}$= 18.23 Mpc/h
but the  averaged radius of the cosmic voids
approximated by spheres is  $\overline {R}$= 22.6 Mpc/h.
The previous  formula contains
the Meijer $G$-function which does not have
an easy numerical implementation.
An approximate PDF
for the radius  of the
cell in $V_p(2,3)$ in the PVT case, 
eqn. (\ref{frmeijer})
can be found  in the framework 
of the  generalized gamma PDF, see \cite{evans} , 
\begin{equation}
G(x;a,b,c) =
\frac
{
c{b}^{{\frac {a}{c}}}{x}^{a-1}{{\rm e}^{-b{x}^{c}}}
}
{
\Gamma  \left( {\frac {a}{c}} \right) 
}
\quad .
\end{equation}
The three parameters  $a$ , $b$ and $c$ 
can be found through 
the  Levenberg--Marquardt  method ( subroutine
MRQMIN in \cite{press})
\begin{equation}
f(r,b)_{app}=
\frac
{
 9.291\, \left( {\frac {r}{b}} \right) ^{ 1.77}{{\rm e}^{- 28.35
\, \left( {\frac {r}{b}} \right) ^{ 8.44}}}
}
{
b
}
\quad ,
\end{equation}
and figure   \ref{f14}  
reports the analytical PDF and  the 
approximate PDF.
This  approximate PDF gives  a relative  error
of $0.5 \%$
in the evaluation of the averaged value.

\section{Conclusions}

The BAO are here analyzed  in the framework
of PVT and  NPVT.
A first test  for the PCF  is done on the  2D/3D  vertexes
of PVT. Figures  \ref{f01}
and  \ref{f02} compare
the well known  mathematical  results  with
our simulation.
The PCF of 2D and 3D vertexes in PVT  presents
a first minimum at $r/\bar{R} \approx 1$
in 2D and at  $r/\bar{R} \approx 1.4$    in 3D.
The  applications  to the local  universe
are done in the framework  of galaxies distributed
according  to  an auto-gravitating
medium in respect to the faces
of the irregular Voronoi's polyhedrons,
see the  PDF  \ref{sech2variance}.
We  first tested  the logistic PDF  for a vertical distribution
of galaxies in 2D/3D    in order to reproduce the observed
small scale behavior of the PCF,
see Figure \ref{f04} in 2D  and   Figure \ref{f05}
in 3D.
The analysis of the oscillations  of the PCF  in the local
universe are split in  two.
We firstly analyzed  the case of a 2D cut
 or $V_{np} (2,3)$  
 which covers
few voids in the presence of an auto-gravitating medium
with a given value  of variance
 $\sigma=0.8 {\rmn {Mpc}}/h $.
In this  2D cut  or $V_{np} (2,3)$    the PCF presents the first minimum
at     $r \approx  67 {\rmn {Mpc}} /h$  when $\bar{R}$ = 42 Mpc.
The second case  is represented by a 2D cut
 or $V_{np} (2,3)$  
  which
covers $\approx 50$ voids
and the
first minimum
is at    $r \approx  65.17   {\rmn {Mpc}} /h$
when $\bar{R}$ = 54 Mpc/h.
The PCF  turns out
to be similar to   the
behavior of  2dFVL sample
, see Figure 3 in \cite{Martinez2009},
where the  first local minimum is
reached at $\approx$ 65 Mpc/h.
Here we have provided an NPVT environment and
distances of the astrophysical simulations expressed in
Mpc/h.
According to the previous results the distances of the PCF
should be expressed in averaged radius units rather than
Mpc/h .
As a practical example when  $\bar{R}$ = 54 Mpc/h
is used as a calibrating value in our simulation,
the first
local minimum of the PCF should be at 1.2 universal units
according to the data  of  the 2dFVL sample,
see Sec. \ref{seclarge}.

\section*{Acknowledgments}
We thank
Tim Pearson
for a useful discussion on the resolution 
of the figures of the package PGPLOT.


\begin{thebibliography}{42}
\newcommand{\enquote}[1]{``#1''}
\expandafter\ifx\csname natexlab\endcsname\relax\def\natexlab#1{#1}\fi

\bibitem[{{Allen} and {Tildesley}(1987)}]{Allen1987}
{Allen}, M.~P. and {Tildesley}, D.~J. (1987), \textit{Computer simulation of
  liquids}, Oxford, NY: Oxford University Press.

\bibitem[{{Angulo} et~al.(2008){Angulo}, {Baugh}, {Frenk}, and
  {Lacey}}]{Angulo2008}
{Angulo}, R.~E., {Baugh}, C.~M., {Frenk}, C.~S., and {Lacey}, C.~G. (2008),
  \enquote{{The detectability of baryonic acoustic oscillations in future
  galaxy surveys},} \textit{\mnras}, 383, 755--776.

\bibitem[{{Balakrishnan}(1991)}]{Balakrishnan1991handbook}
{Balakrishnan}, N. (1991), \textit{Handbook of the Logistic Distribution}, New
  York: Taylor \& Francis.

\bibitem[{{Bertin}(2000)}]{Bertin2000}
{Bertin}, G. (2000), \textit{{Dynamics of Galaxies}}, Cambridge: {Cambridge
  University Press.}

\bibitem[{{Croton} et~al.(2004){Croton}, {Colless}, {Gazta{\~n}aga}, and
  {Baugh}}]{Croton2004}
{Croton}, D.~J., {Colless}, M., {Gazta{\~n}aga}, E., and {Baugh}, C.~M. (2004),
  \enquote{{The 2dF Galaxy Redshift Survey: voids and hierarchical scaling
  models},} \textit{\mnras}, 352, 828--836.

\bibitem[{{Eisenstein} et~al.(2005){Eisenstein}, {Zehavi}, {Hogg},
  {Scoccimarro}, {Blanton}, and {Nichol}}]{Eisenstein2005}
{Eisenstein}, D.~J., {Zehavi}, I., {Hogg}, D.~W., {Scoccimarro}, R., {Blanton},
  M.~R., and {Nichol}, R.~C. e.~a. (2005), \enquote{{Detection of the Baryon
  Acoustic Peak in the Large-Scale Correlation Function of SDSS Luminous Red
  Galaxies},} \textit{\apj}, 633, 560--574.

\bibitem[{{Evans} et~al.(2000){Evans}, {Hastings}, and {Peacock}}]{evans}
{Evans}, M., {Hastings}, N., and {Peacock}, B. (2000), \textit{Statistical
  Distributions - third edition}, New York: John Wiley \& Sons Inc.

\bibitem[{{{Hansen}, J.~P. and {McDonald}, I.~R.}(1986)}]{Hansen1986}
{{Hansen}, J.~P. and {McDonald}, I.~R.} (1986), \textit{{Theory of simple
  liquids}}, New York: {Academic Press}.

\bibitem[{{Heinrich} and {Muche}(2008)}]{Heinrich2008}
{Heinrich}, L. and {Muche}, L. (2008), \enquote{{Second-order properties of the
  point process of nodes in a stationary Voronoi tessellation.}} \textit{Math.
  Nachr.}, 281, 350--375.

\bibitem[{{Hubble}(1929)}]{Hubble1929}
{Hubble}, E. (1929), \enquote{{A Relation between Distance and Radial Velocity
  among Extra-Galactic Nebulae},} \textit{Proceedings of the National Academy
  of Science}, 15, 168--173.

\bibitem[{{Johnson} et~al.(1995){Johnson}, {Kotz}, and
  {Balakrishnan}}]{univariate2}
{Johnson}, N.~L., {Kotz}, S., and {Balakrishnan}, N. (1995),
  \textit{{Continuous univariate distributions. Vol. 2. 2nd ed.}}, New York:
  {Wiley }.

\bibitem[{{Kiang}(1966)}]{kiang}
{Kiang}, T. (1966), \enquote{{Random Fragmentation in Two and Three
  Dimensions},} \textit{\za}, 64, 433--439.

\bibitem[{{Mart{\'{\i}}nez} et~al.(2009){Mart{\'{\i}}nez}, {Arnalte-Mur},
  {Saar}, {de la Cruz}, {Pons-Border{\'{\i}}a}, {Paredes},
  {Fern{\'a}ndez-Soto}, and {Tempel}}]{Martinez2009}
{Mart{\'{\i}}nez}, V.~J., {Arnalte-Mur}, P., {Saar}, E., {de la Cruz}, P.,
  {Pons-Border{\'{\i}}a}, M.~J., {Paredes}, S., {Fern{\'a}ndez-Soto}, A., and
  {Tempel}, E. (2009), \enquote{{Reliability of the Detection of the Baryon
  Acoustic Peak},} \textit{\apjl}, 696, L93--L97.

\bibitem[{{McQuarrie}(1976)}]{McQuarrie1976}
{McQuarrie}, D.-A. (1976), \textit{Statistical Mechanics}, New York: Harper and
  Row.

\bibitem[{{Mehta} et~al.(2011){Mehta}, {Seo}, {Eckel}, {Eisenstein},
  {Metchnik}, {Pinto}, and {Xu}}]{Mehta2011}
{Mehta}, K.~T., {Seo}, H.-J., {Eckel}, J., {Eisenstein}, D.~J., {Metchnik}, M.,
  {Pinto}, P., and {Xu}, X. (2011), \enquote{{Galaxy Bias and Its Effects on
  the Baryon Acoustic Oscillation Measurements},} \textit{\apj}, 734, 94.

\bibitem[{Meijer(1936)}]{Meijer1936}
Meijer, C. (1936), \enquote{{\"Uber Whittakersche bzw. Besselsche Funktionen
  und deren Produkte.}} \textit{Nieuw Arch. Wiskd.}, 18, 10--39.

\bibitem[{Meijer(1941)}]{Meijer1941}
--- (1941), \enquote{{Multiplikationstheoreme f\"ur die Funktion
  $G_{p,q}^{m,n}(z)$.}} \textit{Proc. Akad. Wet. Amsterdam}, 44, 1062--1070.

\bibitem[{{Nadathur} and {Hotchkiss}(2013)}]{Nadathur2013}
{Nadathur}, S. and {Hotchkiss}, S. (2013), \enquote{{A self-consistent public
  catalogue of voids and superclusters in the SDSS Data Release 7 galaxy
  surveys},} \textit{ArXiv e-prints}.

\bibitem[{{Okabe} et~al.(1992){Okabe}, {Boots}, and {Sugihara}}]{okabe}
{Okabe}, A., {Boots}, B., and {Sugihara}, K. (1992), \textit{{Spatial
  tessellations. Concepts and Applications of Voronoi diagrams}}, {Chichester,
  New York}: {Wiley}.

\bibitem[{{Okabe} et~al.(2000){Okabe}, {Boots}, {Sugihara}, and
  {Chiu}}]{Okabe2000}
{Okabe}, A., {Boots}, B., {Sugihara}, K., and {Chiu}, S. (2000),
  \textit{{Spatial tessellations. Concepts and Applications of Voronoi
  diagrams, 2nd ed.}}, {Chichester, New York}: {Wiley}.

\bibitem[{Olver et~al.(2010)Olver, Lozier, Boisvert, and Clark}]{NIST2010}
Olver, F. W. J.~e., Lozier, D. W.~e., Boisvert, R. F.~e., and Clark, C. W.~e.
  (2010), \textit{{NIST handbook of mathematical functions.}}, Cambridge:
  {Cambridge University Press. }.

\bibitem[{{Padmanabhan}(2002)}]{Padmanabhan_III_2002}
{Padmanabhan}, P. (2002), \textit{{Theoretical astrophysics. Vol. III: Galaxies
  and Cosmology}}, {Cambridge, MA}: {Cambridge University Press}.

\bibitem[{{Pan} et~al.(2011){Pan}, {Vogeley}, {Hoyle}, {Choi}, and
  {Park}}]{Vogeley2011}
{Pan}, D.~C., {Vogeley}, M.~S., {Hoyle}, F., {Choi}, Y.-Y., and {Park}, C.
  (2011), \enquote{{Cosmic Voids in Sloan Digital Sky Survey Data Release 7},}
  \textit{ArXiv e-prints:1103.4156}.

\bibitem[{{Press} et~al.(1992){Press}, {Teukolsky}, {Vetterling}, and
  {Flannery}}]{press}
{Press}, W.~H., {Teukolsky}, S.~A., {Vetterling}, W.~T., and {Flannery}, B.~P.
  (1992), \textit{{Numerical Recipes in FORTRAN. The Art of Scientific
  Computing}}, Cambridge: Cambridge University Press.

\bibitem[{{Rohlfs}(1977)}]{Rohlfs1977}
{Rohlfs}, K. (ed.) (1977), \textit{{Lectures on density wave theory}}, vol.~69
  of \textit{Lecture Notes in Physics, Berlin Springer Verlag}.

\bibitem[{{S{\'a}nchez} et~al.(2013){S{\'a}nchez}, {Alonso}, {S{\'a}nchez},
  {Garc{\'{\i}}a-Bellido}, and {Sevilla}}]{Sanchez2013}
{S{\'a}nchez}, E., {Alonso}, D., {S{\'a}nchez}, F.~J., {Garc{\'{\i}}a-Bellido},
  J., and {Sevilla}, I. (2013), \enquote{{Precise measurement of the radial
  baryon acoustic oscillation scales in galaxy redshift surveys},}
  \textit{\mnras}, 434, 2008--2019.

\bibitem[{{Seo} and {Eisenstein}(2007)}]{Seo2007}
{Seo}, H.-J. and {Eisenstein}, D.~J. (2007), \enquote{{Improved Forecasts for
  the Baryon Acoustic Oscillations and Cosmological Distance Scale},}
  \textit{\apj}, 665, 14--24.

\bibitem[{{Seo} et~al.(2012){Seo}, {Ho}, {White}, {Cuesta}, {Ross}, and
  {Saito}}]{seo2012}
{Seo}, H.-J., {Ho}, S., {White}, M., {Cuesta}, A.~J., {Ross}, A.~J., and
  {Saito}, S. (2012), \enquote{{Acoustic Scale from the Angular Power Spectra
  of SDSS-III DR8 Photometric Luminous Galaxies},} \textit{\apj}, 761, 13.

\bibitem[{{Spitzer}(1942)}]{Spitzer1942}
{Spitzer}, Jr., L. (1942), \enquote{{The Dynamics of the Interstellar Medium.
  III. Galactic Distribution.}} \textit{\apj}, 95, 329.

\bibitem[{{Stoyan} and {Stoyan}(1990)}]{Stoyan1990}
{Stoyan}, D. and {Stoyan}, H. (1990), \enquote{{Exploratory data analysis for
  planar tessellations: Structural analysis and point process methods.}}
  \textit{Appl. Stochastic Models Data Anal.}, 6, 13--25.

\bibitem[{{Sutter} et~al.(2012){Sutter}, {Lavaux}, {Wandelt}, and
  {Weinberg}}]{Sutter2012}
{Sutter}, P.~M., {Lavaux}, G., {Wandelt}, B.~D., and {Weinberg}, D.~H. (2012),
  \enquote{{A Public Void Catalog from the SDSS DR7 Galaxy Redshift Surveys
  Based on the Watershed Transform},} \textit{\apj}, 761, 44.

\bibitem[{{Sutter} et~al.(2013){Sutter}, {Lavaux}, {Wandelt}, and
  {Weinberg}}]{Sutter2013}
--- (2013), \enquote{{A response to arXiv:1310.2791: A self-consistent public
  catalogue of voids and superclusters in the SDSS Data Release 7 galaxy
  surveys},} \textit{ArXiv e-prints,1310.5067}.

\bibitem[{{Szalay} et~al.(1993){Szalay}, {Broadhurst}, {Ellman}, {Koo}, and
  {Ellis}}]{Szalay1993}
{Szalay}, A.~S., {Broadhurst}, T.~J., {Ellman}, N., {Koo}, D.~C., and {Ellis},
  R.~S. (1993), \enquote{{Redshift Survey with Multiple Pencil Beams at the
  Galactic Poles},} \textit{Proceedings of the National Academy of Science},
  90, 4853--4858.

\bibitem[{{Totsuji} and {Kihara}(1969)}]{Totsuji1969}
{Totsuji}, H. and {Kihara}, T. (1969), \enquote{{The Correlation Function for
  the Distribution of Galaxies},} \textit{\pasj}, 21, 221.

\bibitem[{{White} et~al.(2010){White}, {Pope}, {Carlson}, {Heitmann}, {Habib},
  {Fasel}, {Daniel}, and {Lukic}}]{White2010}
{White}, M., {Pope}, A., {Carlson}, J., {Heitmann}, K., {Habib}, S., {Fasel},
  P., {Daniel}, D., and {Lukic}, Z. (2010), \enquote{{Particle Mesh Simulations
  of the Ly{$\alpha$} Forest and the Signature of Baryon Acoustic Oscillations
  in the Intergalactic Medium},} \textit{\apj}, 713, 383--393.

\bibitem[{Yarnell et~al.(1973)Yarnell, Katz, Wenzel, and Koenig}]{Yarnell1973}
Yarnell, J.~L., Katz, M.~J., Wenzel, R.~G., and Koenig, S.~H. (1973),
  \enquote{Structure Factor and Radial Distribution Function for Liquid Argon
  at 85 degree K,} \textit{Phys. Rev. A}, 7, 2130--2144.

\bibitem[{{Zaninetti}(2009)}]{Zaninetti2009c}
{Zaninetti}, L. (2009), \enquote{{Poissonian and non-Poissonian Voronoi
  diagrams with application to the aggregation of molecules},} \textit{\pla},
  373, 3223--3229.

\bibitem[{{Zaninetti}(2010)}]{Zaninetti2010a}
--- (2010), \enquote{{ A geometrical model for the catalogs of galaxies },}
  \textit{Revista Mexicana de Astronomia y Astrofisica}, 46, 115--134.

\bibitem[{{Zaninetti}(2012)}]{Zaninetti2012e}
--- (2012), \enquote{{New Analytical Results for Poissonian and non-Poissonian
  Statistics of Cosmic Voids},} \textit{Revista Mexicana de Astronomia y
  Astrofisica}, 48, 209--222.

\bibitem[{{Zaninetti}(2013)}]{Zaninetti2013b}
--- (2013), \enquote{{ Chord distribution along a line in the local Universe
  },} \textit{Revista Mexicana de Astronomia y Astrofisica}, 49, 117--126.

\bibitem[{{Zehavi} et~al.(2004){Zehavi}, {Weinberg}, {Zheng}, {Berlind},
  {Frieman}, and {Scoccimarro}}]{Zehavi_2004}
{Zehavi}, I., {Weinberg}, D.~H., {Zheng}, Z., {Berlind}, A.~A., {Frieman},
  J.~A., and {Scoccimarro}, R. (2004), \enquote{{On Departures from a Power Law
  in the Galaxy Correlation Function},} \textit{\apj}, 608, 16--24.

\bibitem[{{Zhan} and {Knox}(2006)}]{Zhan2006}
{Zhan}, H. and {Knox}, L. (2006), \enquote{{Baryon Oscillations and Consistency
  Tests for Photometrically Determined Redshifts of Very Faint Galaxies},}
  \textit{\apj}, 644, 663--670.

\end{thebibliography}

\end{document}